RESEARCH

# Illustrating Polymerization using Three-level Model Fusion

Ivan Kolesar[1*], Julius Parulek[1], Ivan Viola[1,2], Stefan Bruckner[1], Anne-Kristin Stavrum[1] and Helwig Hauser[1]

*Correspondence:
Ivan.Kolesar@UiB.no
[1] Department of Informatics, University of Bergen, N-5020 Bergen, Norway
Full list of author information is available at the end of the article

**Abstract**

Research in cell biology is steadily contributing new knowledge about many different aspects of physiological processes like polymerization, both with respect to the involved molecular structures as well as their related function. Illustrations of the spatio-temporal development of such processes are not only used in biomedical education, but also can serve scientists as an additional platform for in-silico experiments. In this paper, we contribute a new, three-level modeling approach to illustrate physiological processes from the class of polymerization at different time scales. We integrate physical and empirical modeling, according to which approach suits the different involved levels of detail best, and we additionally enable a simple form of interactive steering while the process is illustrated. We demonstrate the suitability of our approach in the context of several polymerization processes and report from a first evaluation with domain experts.

## Introduction

Polymers are macromolecules that are composed of many smaller molecules, known as monomers. Polymers with different structure and monomer composition have a broad range of different physical properties, like solution viscosity, melt viscosity, solubility, stiffness, and more. Well-known examples of polymers are proteins and the DNA, which play important roles in everyday life. Polymerization is the biochemical process of polymer formation. During polymerization, monomers react with each other to form a macromolecular structure. As polymers are essential components of biological processes, polymerization occurs constantly within the cells of every living organism.

Even though major advances in recent biological and biochemical research greatly extend our knowledge about polymerization, still much remains unknown. With respect to the involved molecular structures, for example, not all of them have been crystallized to derive a better understanding of their spatial structure. Also much remains unknown regarding their physiological function. This naturally inherent uncertainty is one important reason for why it is challenging, for students as well as for professionals from different fields, to form an appropriate mental model of physiological processes.

In order to effectively communicate such processes, it is essential to consider both their spatial and temporal characteristics as well as their multi-scale nature.

Polymerization, for example, ranges spatially from molecules to macromolecules and temporally from nanoseconds (monomer movement) to seconds (overall process of polymerization). It is also not feasible to model the entire physiological processes by just considering the principal laws of physics on the atomic level – we need different models at different levels of details. Moreover, the process of polymerization strongly depends on the properties of the environment such as the concentration of the reacting substances.

In recent years, we have seen a growing number of artistic illustrations of various aspects of cell biology [1, 2] and we have also observed some selected efforts to, at least partially, support the usually cumbersome, manual illustration process with computational tools. However, for a better understanding and for a more effective communication of physiological processes, visualization in the form of static images or animations is often not enough. One should, for example, see the dependence of such a process on its environment and experiment with the interactions between the process and its environment. How will the structure emerge if there are not enough building substances? How do spatial constrains influence branching patterns? An interactive system capable of answering such questions can greatly help to comprehend the process of polymerization and even be an environment for generating or even testing new hypotheses.

For answering the above mentioned questions, a suitable modeling and visualization approach for the interactive illustration of polymerization should satisfy the following requirements:

- It needs to **capture emergence**, i.e., it should be capable of representing the overall process of emergence and its sub-processes, for example, the binding of monomers and branching.
- It needs to represent the **temporal development**, i.e., it has to communicate the time-dependent and dynamic nature of the process.
- The **multi-scale** nature of the process needs to be captured in both space and time.
- **Interactivity** is essential and the user should be able to modify the environment and immediately see the results.
- Even if based on empirical modeling approaches, the illustration must be **sufficiently biochemically correct**.

In this paper, we present a new, three-level modeling and visualization approach, which fulfills the above described requirements. A starting point for our research was the observation that polymerization is physiologically characterized by biochemical processes at different time scales (from nanoseconds to seconds) and that we were aiming at an approach which should be truthful to these different time scales.

The smallest time scale, which we intended to capture with our approach, is the one which corresponds to the diffusion-based movement of monomers nearby the active end of a polymer and the growing of the polymer due to individual monomers that bind to the polymer. Considering the advantages and disadvantages of different modeling approaches (as discussed in the related work section), we concluded that an agent-based system would be best suited to capture the stochastic characteristic of the movement of the monomers.

On the other end of the temporal scale space, we intended to capture the entire growth process of a polymer – a process which is many orders of magnitudes slower than the diffusion-based movement of the monomers. We understand that these polymerization processes (at a larger time scale) are much more deterministic in terms of their development. Therefore, it is appropriate to model this process at this level by means of an L-system (this is also in line with many other cases of biological growth, like plant growth [3, 4, 5], which regularly are modeled the same way).

To realize a solution which is capable of representing both of these physiological aspects, we devised an approach which integrates both modeling concepts. We find it reasonably straight-forward to formulate rules for an L-system so that it models the overall growth of a polymer. We link – via a communication system (see the Communication & Process specification section for more details) – the agent-based system to the L-system so that certain rewriting rules of the L-system – in particular those, which correspond to the binding of a monomer to the polymer – only complete, if they are supported by the linked agent-based system.

Furthermore, we intended to also enable a minimum amount of interactive steering – at least to the degree that the user can influence the environmental conditions of the polymerization process to a certain degree. To achieve this, we couple the agent-based system with another modeling layer, i.e., a density-based modeling layer (here called "system of densities", SOD). On this layer, we only consider the overall densities of all involved building blocks (mostly the monomers) and we maintain a process – parallel to the overall modeling process – which, at any time, influences the agent-based system so that the number of agents in the multi-agent system corresponds, as good as possible, with the corresponding densities in the SOD. By interactively modifying selected densities in the SOD, the user can thereby, to a certain degree, steer the polymerization process.

After we first discuss related work in the following, we then go into more technical details with respect to our solution. We also report from an evaluation which we conducted together with several domain experts.

## Related work

As mentioned above, our work is based on a fusion of three different modeling techniques, i.e., an L-system, an agent-based system, and a system of densities. In the following, we comment on the state of the art with respect to all of these individual approaches, as well as on previous attempts to extend them.

### L-Systems

Lindenmayer systems [6] are a broadly used modeling approach for the development of linear and branching structures, built from discrete modules. An L-system can be seen as a formal, parallel rewriting grammar. It consists of an alphabet of symbols, a collection of rules that expand symbols into new symbols, or strings of symbols, an initial string, called axiom, and a mechanism for translating the generated string into an according geometric structure. Since the introduction of L-systems in the late 1960s, many extensions to the original approach were proposed, such as *stochastic*, *context-sensitive* and *parametric* L-systems, many of which are well described in a book by Prusinkiewicz and Lindenmayer [3].

Originally, L-systems lacked one important aspect of structural modeling, which is the interaction between the structure and its environment. The first extension that related L-systems to an environment as an affecting factor, were *parametric* L-systems [7, 3]. Here, every symbol is extended by its own parameter space, that is applied and changed by the production rules.

An *environmentally-sensitive* L-system [8] contains local, rather than global, properties of the environment that affect the model. This concept is based on query symbols, which return the position and orientation of the current, graphically interpreted symbol, in the given coordinate system. These parameters are passed as arguments to user-defined functions which then return local properties of the environment for the inquired location.

A more general approach for the communication between the model and the environment was introduced in the *open* L-systems [4]. This technique extends environmentally-sensitive L-systems by using a special symbol for bilateral communication with the environment. The environment is no longer represented as a simple function, but becomes an active process that may react to the information from the model. Open L-systems were used for modeling the development of different structures such as ecosystems [9, 4], cities [10], proteins folding [11], plants, trees and roots [5, 12], or even fire [13].

In our case, we find L-systems only partially suitable. While we, on the one hand, find them useful to represent the large-scale aspects of polymerization, their utility is, on the other hand, also limited, since they cannot intrinsically capture crucial small-scale characteristics of polymerization – in particular, the interaction of many individual actors (most importantly, the monomers and their behavior). Strengths and weaknesses of L-systems, with respect to modeling an illustration of polymerization, are shown in Table 1.

### Agent-based systems

In contrast to L-systems, agent-based modeling [14] is centered around multiple autonomous entities called agents. Agents are computing elements with two important capabilities [15]. Firstly, they are capable of autonomous action, i.e., they can act independently in order to satisfy their designed objectives. Secondly, they are capable of interacting with other agents. An agent's behavior is defined to achieve an individual or collective objective.

This modeling approach provides a natural metaphor for understanding and building a wide range of systems, such as social systems, biological systems, economics, traffic or transportation systems that feature many independent actors which drive the system's global behavior.

In the context of emergent phenomena, agent-based systems have been employed in modeling molecular self-assembly [16, 17] and intracellular interactions [18, 19].

As agent-based systems model a global behavior through the interaction of individual entities, they are well suited for the purpose of modeling the crowded environment of the cell. However, a major drawback is that the global effect resulting from the interaction of the individual agents is very difficult to control and steer. In our case, we find agent-based modeling suitable for the small scale of polymerization, i.e., the movement of the monomers, etc., while we require better control over the modeling when considering the process on a larger scale.

### Integrated approaches

As shown in Table 1 both L-systems and agent-based modeling have strengths and weaknesses. Naturally, one thinks about the combination of both concepts to get the advantages of both approaches while mitigating their disadvantages. One way of integrating both approaches, researched by von Mammen, is *swarm grammars* [20, 21]. Swarm grammar was developed as an integrated representation of artificial crowds and a developmental model. In this approach, the L-system doesn't hold the information about a structure, but about the agent's states in the environment and is the deterministic tool for the evolution of the agents over time. The usefulness of such an approach was exemplified in the generation of the 3D geometry from the agents state's [22] and the application of this method to architectural design [23]. However, with this modeling approach the graphical representation describes the development of the crowd, not the development of the structure. Moreover this approach doesn't provide the modeling solution for a bidirectional communication between the structure and the agents and is therefore not suitable for the interactive illustration of polymerization.

Other modeling approaches are based on the combination of rule-based and particle-based reaction and diffusion modeling [24, 25]. In these approaches the resulting molecular structures are represented as a graph, where each node is an elementary unit, for example, a simple molecule or a monomer. The molecules are defined as spatial particles and their behavior in the environment is described by molecular dynamics and reaction rules. The resulting molecule is a stochastically built predetermined structure. These modeling approaches are using different visualization software (SRSim [24], ZygCell3D [26]), which provides direct visualization of the modeled polymerization.

In our modeling approach, we are introducing the probabilistic variability, i.e., the resulting molecular structure is not predetermined. With the L-system, our approach is capable to represent not only the information about the current structure, but also the information about processes that are currently associated with it. Furthermore, we know that the time scales between the overall process of the creation of the structure (seconds) and the movement of the single independent molecule in the environment (nanosecond) are largely different. We address these time scale differences by the possibility to interactively change the current time scale and the ability to switch between them. This helps to comprehend the creation of the structure and the relation to the different time scales of the process. Also for experiments, our solution provides steering of the simulation by changing the density (concentration) of the molecules in the environment. On top of that our solution provides a tool for changing the rules that define processes (reactions) during the simulation. Our approach provides a direct 3D visualization of the processes, but we can easily encode additional information in the visualized structure, for example the uncertainty of the creation of branches.

## Method overview

Our solution is composed of several different sub-systems (see Figure 1), which are mutually synchronized with each other. The simulation runs in a cuboid domain of changeable dimensions with a time step of length $\Delta t$.

**Algorithm 1** Overall simulation

```
1:  t ⇐ 0
2:  Δt ⇐ deltatime
3:  AS ⇐ intialize(AS)
4:  CS ⇐ intialize(CS)
5:  LS ⇐ intialize(LS)
6:  SOD ⇐ intialize(SOD)
7:  while running do
8:      Eval(LS)
9:      Visualize(LS)
10:     Eval(SOD)
11:     Eval(CS)
12:     if P(Δt) < timeScaleTresh then
13:         Eval(AS)
14:         Visualize(AS)
15:     end if
16:     t ⇐ t + Δt
17: end while
```

As depicted in Algorithm 1, the simulation starts with setting the simulation time $t$, the current delta time of the simulation $\Delta t$ and initializing the systems of the simulation: L-system (LS), communication system (CS), agent-based system (AS) and system of densities (SOD). The basic cycle, shown also in Figure 2, is composed of the following steps:

a) The L-system is evaluated, which involves processing the communication with the monomers and growing the polymer if a new monomer binds to the growing end. (Line 8)
b) The L-system structure is visualized. (Line 9)
c) The SOD verifies the current densities and communicates the required changes to the agent-based system. (Line 10)
d) The communication system firstly evaluates on which time scale is the simulation currently running. This is done by the evaluation of the function $P(\Delta t)$, which is described in more detail in the Communication & Process specification subsection. If $P(\Delta t) < timeScaleTresh$, i.e., the time delta is relevant for monomer motion, the communication system transfers communication parameters from the L-system to the agent-based system and vice versa. In the case that $P(\Delta t) > timeScaleTresh$, the growth is computed from the probability of function $P(\Delta t)$. (Line 11)
e) If $P(\Delta t) < timeScaleTresh$, meaning the simulation is in the monomer motion time scale, the agent-based system is evaluated and visualized. (Lines 12, 13, 14)

In the following subsections we provide a more detailed description of the mentioned components.

## L-System

The L-system consists of an ordered triplet $L = \langle A, \omega, P \rangle$, where $A$ denotes an alphabet, $\omega$ is a non-empty word called axiom and $P$ is a finite set of production rules. The axiom $\omega = (a_i, a_i \in A)_{i=0}^{n}$ defines the initial development of a polymer of size $n$ in the simulation.

The symbols of the alphabet $A$ are divided into four semantic categories: *Binding*, *Structure*, *End*, and *Communication* symbols. A *Structure* symbol represents a monomer and holds information about the monomer type and its geometry. A *Binding* symbol represents the binding relation between two monomers and holds information whether the binding point is a start of the new branch. The end of a branch is encoded by the *End* symbol. These symbols describe the structural aspects of a polymer in the L-system.

Processes are represented by *Communication* symbols. A communication symbol has the role of a bi-directional bridge between the L-system and the agent-based system through the communication system. It is defined by $C(O, Type, t, r)$, where $O$ identifies the process, e.g., growing or branching, $Type$ is the identification of the agent type the process is connected to, for example, $t$ is the process lifespan and $r$ encodes the result of the process. For example, the communication symbol $C(binding, glucose, 5.0, r)$ queries information about the process binding the glucose molecule and expects the result in parameter $r$. Communication symbols have a global parameter $t_{max}$ defining the maximum allowed time that the process can take. If the process is about to take longer, it is terminated.

A production rule from $P$ has the following format [4]:

$$id : predecessor : condition \longrightarrow successor : probability$$

where *id* is the rule identifier (label), *predecessor* is a symbol that will be replaced by the *successor* symbol, but only if *condition* is evaluated as *true*. The *probability* part represents a chance value that this production rule will happen at all.

The L-system has two important phases: derivation and interpretation. The derivation step is the rewriting process: $\omega_i \xrightarrow{P} \omega_{i+1}$. In each step, the production rules $P$ replace all predecessor symbols $\omega_i$ by successor symbols, generating a new string $\omega_{i+1}$.

The derivation step is followed by an interpretation step that transforms a string of symbols into a 3D geometrical representation. During the interpretation step, the string is read from left to right by an interpreter. The interpreter stores its spatial position $I_{pos}$ (vector) and orientation $I_{ori}$ (quaternion). These variables are initialized at the beginning of the interpretation step by the position and orientation of polymer starting point. When the interpreter reads a structure symbol, then it places the geometry specified by it into the scene according to the current $I_{pos}$ and $I_{ori}$. When the interpreter reads a binding symbol, it updates its position and orientation as follows:

$$I_{pos} = I_{pos} + I_{ori}Bin_{pos}$$
$$I_{ori} = I_{ori}Bin_{ori},$$

where $Bin_{pos}$ and $Bin_{ori}$ are the binding position (vector) and orientation (quaternion) from the binding symbol. By this transformation the system can create the geometric representation of the whole polymer (Figure 3). Also, during this interpretation step the position and orientation parameters of the communication symbols are updated with the $I_{pos}$ and $I_{ori}$ of the current state.

Essentially, the evaluation of the L-system depicts the development of the polymer growth. First, the $r$ parameters of the communication symbols are filled with values, retrieved from the communication system. Next, the derivation and interpretation phases are applied.

For example, let us define an L-system with the axiom $C(grow, molecule, 0, \varnothing)$ and the following production rules:

$$p_1 : C(grow, molecule, t, r) : r \neq \varnothing \longrightarrow mC(grow, molecule, 0, \varnothing)$$
$$p_2 : C(grow, molecule, t, r) : t > t_{max} \longrightarrow \varepsilon$$

The $t_{max}$ parameter is an empirically chosen time limitation of the *grow* process. In the beginning of the L-system evaluation the $t$ and $r$ parameters of the C symbol are retrieved from the communication system. Afterwards, in the derivation phase, the production rules are applied.

Only the rules with the same predecessor and correct predecessor parameters are applied. For example, in a case when $t = 0.05$ and $r = \varnothing$, during the derivation step no production rules can be applied since both conditions $r \neq \varnothing$ and $t > t_{max}$ of the rules $p_1$ and $p_2$ are not met. In this case, the L-system's string is left unchanged.

When the agent system, through the communication system, returns values $t = 0.05$ and $r = molecule$, the derivation step applies rule $p_1$ and produces the new string $\omega = mC(grow, molecule, 0, \varnothing)$ with a new symbol $m$, and the communication symbol is replaced by $C(grow, molecule, 0, \varnothing)$. This means that the growing process has finished and a new process of growing is created at the end of the structure.

If the process takes too long for values $t = 5.05$ and $r = \varnothing$, rule $p_2$ is applied, rewriting the communication symbol to the end symbol; i.e., the growing process of the current branch is terminated.

### Communication & Process specification

The information exchange between the L-system and the agent-based system is realized through the communication system. The behavior of this system depends on the current time scale of the simulation.

If the simulation is running in the time scale of monomer motion, the communication system retrieves the processes parameters from the L-system and transports them in a form of queries to the agent-based system. After the simulation step of the agent-based system, the communication system retrieves the results of the agent-based system queries and feeds the results to the communication symbol of the L-system.

The queries are represented as a $Q(pos, ori, type, time, result)$. The position, orientation and type parameters are retrieved from the L-system interpreter; and copied into *pos*, *ori* and *type*. The agent-based system updates the parameters *time* and *result*. The *result* is an agent type and the system fills this value if and only if an agent of the specified type reaches the position *pos* with the orientation *ori*.

On the other hand, if the simulation runs on the time scale of the whole process, the agent-based system does not participate in the communication. Instead, the communication system applies the function $P(\Delta t)$, computing a probability of the temporal event to the *result* of query $Q$. The function $P(\Delta t)$ is a probabilistic description of the process with respect to $\Delta t$. An example of this function is shown in Figure 5. Function $P$ should return 0 if the $\Delta t$ is lower then the threshold for time scale switching, and a value from 0 to 1 for a larger $\Delta t$. The assignment of the agent-based system and $P(\Delta t)$ to the *result* parameter is described by the following equation:

$$R(\Delta t, t) = P(\Delta t)d_{type}(t)a_{type} + (1 - P(\Delta t))AS(t),$$

where the function $P(\Delta t)$ is the aforementioned probability function. The first half of the function $P(\Delta t)d_{type}(t)a_{type}$ denotes the return value if the simulation happens at a larger time scale. The second part of equation, $(1 - P(\Delta t))AS(t)$, applies the return value from the agent-based system $AS(t)$ in the lower time scale.

Importantly, the global parameter $\Delta t$, together with the description of the process behavior $P(\Delta t)$, can be interactively changed. This interactivity enables us to model and visualize polymerization processes across different time scales during the simulation.

### Agent-based system

An agent-based system is utilized to capture the stochastic motion characteristics of monomers and the binding processes. The agent-based system is defined as $AS(t) = \{a, b, c, ...\}$ where $t$ is a global time parameter and $a, b, c, ...$ are sets of different types, in our case molecules.

Each agent has the following attributes: position, orientation, velocity, angular velocity and type. Additionally we define a set of functions representing its *conditions*, *behaviors* and *triggers*. Behaviors define the agent actions, conditions constrain agents within spatial boundaries and triggers are functions that are conditionally executed. The behavior of agents are not limited only to physical behavior. In our agent-based system the behavior of the agents can be defined to generally illustrate the process or to realistically simulate the required behavior.

In our case we wanted to have illustrative the diffusion movement and the binding process. However, there is a large time scale difference between them. The diffusion movement of the molecules are much faster than the binding process. Moreover, the time distance, in the time scale of binding, between two binding processes is comparably large. Therefore the agent-based system applies two types of approximation to the monomer movement depending on if the aim of visualizing is the movement of monomers or the binding process.

If the agent-based system is used to interactively visualize the binding process of a monomer, random walking is applied to approximate the diffusion [27]. The agent position is updated by a random velocity vector with a diffusion coefficient at every time step. It would take a long time if we would stay in this time scale and wait for a new molecule to come to the binding site and bind. Therefore if there is no binding process to illustrate, the simulation is fast-forwards to the next binding process. During this stage the molecules are moving so quickly, that there is no visual correlation of monomers between two time steps. In this stage the monomers' position and orientation are calculated based on a random distribution.

It is important to point out, that our aim is to sufficiently correctly illustrate the effect of diffusion and binding, not to realistically reproduce it. The speed of the process of monomer binding can be interactively altered by the global parameter $\Delta t$ that specifies the amount of time between two simulation steps.

### System of Densities

Here, we consider the overall densities of all involved agents of the agent-based system. The SOD is defined as a set of functions $SOD = \{d_a, d_b, d_c, ...\}$. Each function represents the density of an agent type over time.

Parallel to the other models, in every time step the SOD attempts to keep the number of agents $\|a\|$ as close as possible to $d_a(t) \times V$, where $V$ is the volume of the space in which the agents simulation run. The user can steer the polymerization interactively by modifying the densities in the SOD. Figure 4 illustrates the behavior of the steering option.

## Implementation

Our implementation is based on the Unity3D framework [28]. This game engine is becoming increasingly popular, also within the bio-community [29]. Its simple $C\#$ programming interface provides fast prototyping possibilities and its efficient plugin system allows quick sharing of results, e.g., utilizing the Unity3D web-plugin.

### Visualization

Our polymerization visualization exploits 2D and 3D features of Unity3D. The number of molecules in both the agent-based system, as agents, and the L-system, as structural symbols, is in the order of thousands.

The geometrical representation of the molecules was generated with the VMD [30] software from PDB files. VMD is developed with NIH support by the Theoretical and Computational Biophysics group at the Beckman Institute, University of Illinois at Urbana-Champaign. The position of binding sites were also gathered from the PDB files and binding orientations were set manually from collected knowledge about the final appearance of the structures.

Each molecular mesh is obtained by means of the solvent excluded surface representation [31], which subsequently was simplified for performance reasons. This is because the generated raw molecular meshes are large (hundreds of thousands of triangles) and cause a performance bottleneck when using them. Thus, we sacrifice some geometric accuracy in order to devote more computational resources to the execution of our model.

We furthermore utilize screen space effects that add illustrative aspects to the eventual rendering (Figure 7). Namely, we perform an outline contour enhancement and screen space ambient occlusion [32].

### Interactivity

It is important to mention that all parameters regarding the shape and the visual molecular appearance can be adjusted by the user in the process of setting up the simulation through the Unity3D GUI.

The Unity3D GUI was also customized for the purposes of interactively changing selected parameters of the system during the simulation. The global system parameter $\Delta t$ controls the speed of monomer movement during the binding process and also serves as an input for the switching between the time scales. The molecules are visualized differently for the different time scales. In the time scale of the binding process, the molecules are opaque. During monomer movement, the molecules are semi-transparent and in the time scale of the overall process the molecules of the agent-based system are hidden.

During the lower time scale of the simulation, the agents count (per type) can be interactively changed (Figure 6) in the SOD. The user can interactively add new types of molecules or alter the concentrations of existing molecules in the environment.

The Unity3D Editor allows even pausing the application and provides the opportunity to change the settings of the system. L-system rules can be added, changed or removed while the simulation is suspended. For example, the user can pause the simulation, and increase the probability of branching of the structure, by increasing the probability of the branching rule and decreasing the probability of the growing rule.

## Examples

Examples of naturally occurring polymers are DNA, proteins, glycogen, starch and poly-ADP-ribose. The structure of polymers is important for their physical properties, for example solubility. This can be exemplified by looking at the properties of glucose polymers. Starch is a carbohydrate used to store energy in plants. It consists of two types of molecules, amylose and amylopectin. Amylose is composed of linear chains of glucose monomers and is insoluble in water, while amylopectin is composed of branched chains of glucose monomers, and is soluble in water. Polymers that contain one type of monomer are referred to as homopolymers, while polymers containing more than one type of monomer are referred to as heteropolymers. The DNA and proteins are made up of four and 20 monomers, respectively, hence are examples of heteropolymers. Glycogen, starch and poly-ADP-ribose are examples of homopolymers.

Here we model reactions of glucose to form cellulose, ADP-ribose to form poly-ADP-ribose, and the creation of microtubules as examples of different types of bio-polymer architecture and composition. The results of our method is shown in Figure 8. Our modeling approach and interactive simulation provides a visual environment for helping users (e.g., students) to understand the processes.

### Cellulose

Cellulose is an important structural component of plant cell walls and is one of the most common organic polymers on the planet [33]. It is made up of long unbranched chains of D-glucose, that are joint together by beta-1,4 glycosidic bonds. The length of the polymers may vary from a few hundred to thousands of monomers. Each D-glucose monomer is rotated 180 degrees compared to the previous monomer in the chain. Parallel chains of cellulose may bind to each other to form secondary structures with various degrees of order. All of this results in fibers with various properties, and much research in the last 100 years have gone into understanding how this can be exploited in the industry.

Cellulose represents an example for the creation of linear homopolymers. In this example, we have molecules of D-glucose floating around in the environment. The polymer, and its creation, is expressed in the L-system with the symbolic alphabet $\alpha = \{m, g, C(growth), \varepsilon\}$. Where $m$ is the structural symbol representing D-glucose, $g$ is the binding symbol specifying that the next structure in the line will be placed above carbon 4 of D-glucose and rotated by 180 degrees. Lastly, $C(growth, Dglucose, t, r)$ is a communication symbol specifying the process of growth by binding a new agent of type D-glucose to the structure with the process time $t$ and the current process result $r$.

The rules from Appendix 1 were used for this example. The first rule $p_1$ dictates, that if the result $r$ of the symbol $C$ is non-empty then the structure is extended by a new subunit $m$ with position and rotation defined by $g$ and on the end of this structure starts a new process of growing $C(growth, Dglucose, 0, \varnothing)$.

The mesh representation of the D-glucose molecule was exported from PDB with the VMD software. An outcome of the modeled cellulose polymerization is shown in the first row of Figure 8, where D-glucose molecules are visualized with green material.

### poly-ADP ribose

ADP-ribose is formed by cleaving Nicotinamide adenine dinucleotide (NAD) to form Nicotinamide and ADP-ribose. The ADP-ribose units may be attached to a variety of proteins, which create various signaling events in a cell [34]. Some of the events are triggered by attaching single ADP-ribose units, while other events are triggered by building ADP-ribose polymers on proteins. One event dependent on ADP-ribose polymers is NAD-dependent DNA repair. Single-strand breakage (SSB) or double-strand breakage (DSB) can potentially be very harmful to a cell unless properly repaired. Poly (ADP-ribose) polymerase (PARP) is an enzyme found in close proximity to the DNA, and is activated by SSB and DSB. It binds to the damaged site to protect the DNA ends, until the repair enzymes are in place. Once attached to the DNA, PARP auto-modifies itself by cleaving NAD molecules and attaching the resulting ADP-ribose monomers to a growing ADP-ribose polymer on itself. The final poly-ADP-ribose structure contains about 200 monomers with about 20-25 monomers per branch. ADP-ribose is negatively charged. This helps to recruit proteins involved in the DNA repair to the site. Since DNA is also negatively charged the growing tree will in addition pull PARP off the DNA, due to electrostatic forces. This makes room for the DNA repair enzymes to come in and repair the damaged site [34].

Poly-ADP-ribose represents the example for the creation of branched homopolymers. In the agent-based simulation, we have agents for NAD and other molecule types. The L-system alphabet $\alpha = \{m, g, b, C(grow), C(branch), \varepsilon\}$ is composed of the structural symbol of ADP-ribose $m$ , binding symbols $g$ and $b$, where $b$ is the beginning of a branch in the structure and $g$ is the continuation of the branch. The communication symbols $C(grow)$ and $C(branch)$ describe the growing and branching processes.

For the polymerization of poly-ADP ribose the production rules from Appendix 2 were used. The development starts with the initial growing process $C(grow, NAD, t, r)$. Rules $p_1$ and $p_2$ are controlling the growth process of the structure and the percentage chance of starting the process of branching. When the branching process is finished, $p_3$ creates the new branch and starts its growing. Rules $p_4$ and $p_5$ are aging rules, saying that if the process is not finished in time $t_{max}$, it will be terminated. The creation of poly-ADP ribose is shown in the second row of Figure 8. The NAD is visualized with red material. As soon as the NAD is processed and as ADP-ribose is composed to the structure, the color of the molecule is changed from red to white. The other molecules in the environment of nucleus are colored with green and blue material.

### Microtubules

Microtubules are long tubular polymers that are involved in a number of important cellular processes. They are found in the cytoplasm of eukaryotic cells, where they acts as part of the structural framework that determines cell shape and cell movements. Microtubules also have important roles in the cell division and act as a railway system for intracellular transport. The microtubule polymers consists of repeating units of a globular protein called tubulin. Tubulin is a dimer which is made up of two polypeptides, called alpha and beta tubulin. A microtubule generally consists of 13 protofilaments [35] assembled around a hollow core. The protofilaments are composed of arrays of tubulin dimers, that are arranged in parallel. The assembly and disassembly of microtubules are highly dynamic. A detailed review of these processes can be found in the review by Akhmanova et al. [36].

From the structural, and content point of view, the microtubule represents an example of linear heteropolymers. For this example, the agent-based system contains agent's types of tubulin and background molecules. The Tubulin agent is composed of coupled agents of alpha tubulin and beta tubulin. The L-system has an alphabet $\alpha = \{a, b, v, h, C(grow)\varepsilon\}$, where $a$ and $b$ are structural symbols of alpha tubulin and beta tubulin. The binding symbols $v$ and $h$ define the binding between the alpha and the beta tubulin, which creates the inner structure of the tubulin dimer, and the binding between two neighboring dimers. The process of growing the structure is described by the communication symbol $C(grow)$.

The corresponding rules from Appendix 3 define the overall microtubule creation. The rule $p_1$ attaches the monomers of the tubulin dimer (alpha and beta tubulin) to the structure and continues the growing at the end of the structure. The third row of Figure 8 shows different stages of the development, where the new dimer is always connected to the end of the spiral. The polymerization of microtubules, as described in [36], is believed to occur in sheets which fold into the circular structure. Our visualization differs from this description (tubular geometry is produced directly) since we do not model the forces necessary to complete the folding process.

The microtubule example is shown in the third row of Figure 8. The tubulin dimer is visualized as couple of alpha tubulin, with light blue, and beta tubulin, with dark blue.

### Synthetic, non-biological showcase

Our approach can model the emergence of more complex structures than what was described in the previous examples. In this example, we demonstrate the creation of complex branching patterns in an overall structure with different types of subuints. Our L-system has an alphabet composed of structural symbols $a, b, c$, binding symbols $a_g, a_b, a_{s1}, a_{s2}, a_{s3}, b_g, c_g$ and communication symbols C(aGrow), C(aBranch), C(aBranchGrow), C(aStar1), C(aStar2), C(aStar3), C(bGrow) and C(cGrow). The Symbols $a$, $b$, $c$ are geometricaly represented as sphere, cube and cylinder. The binding symbols $a_g$, $a_b$, $a_{s1}$, $a_{s2}$ and $a_{s3}$ are the binding sites between spheres in a main branch, main and a secondary branch, main and first star branch, main and second and main and third branch. Lastly, we have communication symbols for subprocesses of growing a main branch C(aGrow), secondary branches from the main branch C(aBranch), their growing C(aBranchGrow), the creation of the star architecture composed of three branches (C(aStar1), C(aStar2), C(aStar3)) and growing of those branches by alternately binding different type of structures (C(bGrow), C(cGrow)).

The L-system rules for the overall process are defined in Appendix 4. Rules $p_1$, $p_2$ and $p_3$ are responsible for the growing of the main branch and starting the processes of creating the branches. The rules $p_4$, $p_5$ set the creation and growth of the branches from the main branch. Lastly the rules $p_6$, $p_7$, $p_8$, $p_9$ manage the creation of the star architecture on the top of the structure, stopping the growth of the main branch. These also manage the growth of the star branches in a way that two types of subunits are placed periodically.

## Evaluation

We have discussed the presented examples of our system with two experts in the field of biology and bioinformatics and one expert from the molecular illustration field. The demonstration of our system was presented as a video showing the animations of the simulations of the mentioned biological examples. Also the interactivity of the system was presented by video to show the changes of the system based on changing the parameters. For every example, we provided the biological explanation and afterwards the users observed the system for several minutes.

Professor Mathias Ziegler, expert in the field of biology, was impressed by the outcome of our approach. He mentioned that the system could generate several proto-structures and model energy requirements for the reactions. With this extension he could imagine that it may be used for the generation (and even for the testing) of hypotheses. For example, one question, to which our system could possibly bring an answer is, what is the ideal branching percentage for the best coupling of glycogen.

He particularly appreciated the system of density layer for the control of the molecule counts during the simulation and the interactive change of modeling rules. In his opinion, the outcome of our work can be used for teaching purposes. Especially, he was impressed by the capability of our system to create complex structures simply from information of the geometrical representation of subunits, their binding sites and simple rules.

Another expert, Assoc. Prof. in Molecular Bioinformatics, suggested that we could show the outcome of our system in the context of examples of complex multimeric structures, especially when it comes to complex formation. Additionally, she pointed out that all polymer formations are catalyzed by enzymes and in many cases this is what determines the later structure as well as the speed of the assembly. With this addition we could provide better biological understanding of these processes in the context of teaching. She also pointed out that with further extensions of the work we could be able to bring answers to some unsolved questions in the field of polymer synthesis. Another aspect in the context of polymerization is the possibility that a local depletion of pre-cursors might be the factor that limits the chain length.

We also discussed our approach with a professional illustrator. She pointed out the importance of having a system for generating a complex, dynamic and accurate biological scene in a timely and financial manner. Being able to easily generate dynamic, accurate and aesthetically pleasing molecular scenes is extremely beneficial for animators and scientific filmmakers.

From a biomedical animation point of view, she praised the system as a quick, easy to use and flexible tool for generating good quality and aesthetically pleasing images. However, she was missing more control over rendering styles and lighting. While she saw the system as an excellent start, being able to bring these dynamic systems directly into 3D animation software would be, in her opinion, ideal. Overall, she considered the biological scenes generated from this system useful for producing biological animations.

Many of the ideas of the domain experts, are good suggestions and should be considered in follow-up future work.

## Discussion and Limitations

Our modeling system is composed of three main parts, i.e., the L-system with communication symbols, an agent-based system and a system of densities. Their behavior and their interactions are determined by defining the agent's behavior, and their numbers and by specifying the L-system's alphabet and production rules.

We demonstrated the use of this modeling system in the context of several examples from molecular biology that capture the creation of different types of polymers. We found out that the proposed modeling and visualization system makes it possible to easily create, modify, and visualize the models on different spatial and temporal scales. The simulations of the polymerization were fast enough to allow interactive experimentation with the models.

In the process of developing this model we became increasingly aware of the lack of information about the creation of polymer structures. This opens a door for the possibility to use our approach for hypothesis generation or at least as a testing environment for the polymerization.

While the main goal of our work was to develop an approach capable of creating interactive educational and illustrative visualizations of biochemical processes, our approach can also help to develop a better understanding of the uncertainties in the underlying model. For instance, as the branching probability has considerable impact on the resulting geometrical structure, it is interesting to explore its influence.

During the development of the structure there is always a chance that either the structure will grow, or create a branch. This chance is defined as probability of executing one of the rules with the same predecessor, where one rule is for growing and another for branching. The closer the rule chance is to this probability, the bigger is the uncertainty of the branching. This is accumulated in the new branches during structure development and encoded as one of parameters of structural symbol. On the other hand, if no branching is applied from the monomer it is interesting to visualize how close was this chance for branching and this is similarly as uncertainty encoded as parameter of structural symbol.

The structural symbol of the L-system is not restricted to the parameters for location and orientation of the molecule and its geometrical representation. Figure 9 provides the visualization of the created structure and newly mentioned parameters of structural symbol: branching uncertainty (white to red) and branching probability (white to blue). Our approach flexibly supports the study of this and similar properties of the model and can therefore has the potential to provide valuable insights beyond the generated geometric structures.

The limitations of the current system version include the absence of modeling third parties in the process, for example enzymes. Additionally, the rules of the L-system are not context-sensitive, meaning that we are unable to model sub-processes, which depend on neighborhood information in the structure. Another challenge is represented by incorporating rigid body simulation and force field to the resulting structure, simulating biologically feasible, dynamic behavior and processes dependent on them, as was pointed out in example of microtubules polymerization.

## Conclusions

We have presented a novel modeling approach that is capable of illustrating the polymer emergence within a filled environment of stochastically moving molecules. Our approach by a complementary fusion of three systems gets strong points from three different modeling approaches. Our approach can model, simulate and interactively visualize the emergence in the stochastic environment in the different time scales. The essential elements of this modeling approach are:

- emergence – model the creation and development of a structure
- temporal development – communicates the time-dependent and dynamic nature of the process
- multi-scale – captures the process on different scales of space and time
- interactivity – a steering tool to influence the environmental conditions

We demonstrated the possibilities of the model in examples of polymerization of linear and branched polymers with one or several types of monomers.

The fusion of models could also be potentially used in other applications, for example to model the emergence of coral reefs, bacterial cultures, or in fields outside of biology, e.g., for the procedural modeling of cities, growth of infrastructure, or emergence of crystals.

**List of abbreviations used**

ADP: Adenosine diphosphate ribose, DNA: Deoxyribonucleic acid, DSB: Double-strand breakage of the DNA, NAD: Nicotinamid adenin dinucleotide, NIH: National Institutes of Health, PARP: poly (ADP-ribose) polymerase, PDB: Protein Data Bank, SOD: System of densities, SSB: Single-strand breakage of the DNA, VMD: Visual Molecular Dynamics

**Competing interests**

The authors declare that they have no competing interests.

**Author's contributions**

Ideas and concept were jointly discussed among all authors. IK developed the model, implemented the prototype and wrote the paper. JP discussed the implementation of visualization part and wrote several parts of paper. IV discussed and brought the idea of time-scale switching method and usage of the L-system for development of polymers. AKS introduced the biological background, consulted the choices with biology scientists and wrote the introduction parts to the biological examples. SB brought the ideas of combination of L-system and agent-based system. HH introduced the third layer of the model - SOD and wrote several parts of paper. All authors read and approved the final manuscript.

**Acknowledgements**

This work has been carried out within the PhysioIllustration research project (# 218023), which is funded by the Norwegian Research Council. This paper has been also supported by the Vienna Science and Technology Fund (WWTF) through project VRG11-010, additionally supported by EC Marie Curie Career Integration Grant through project PCIG13-GA-2013-618680. We would also like to thank our application domain expert in the biology: Mathias Ziegler for his useful ideas and feedback.

**Appendix 1: Cellulose L-system rules**

$$p_1 : C(growth, Dglucose, t, r) : r \neq \varnothing \longrightarrow gmC(growth, Dglucose, 0, \varnothing)$$
$$p_2 : C(growth, Dglucose, t, r) : t > t_{max} \longrightarrow \varepsilon$$

**Appendix 2: poly-ADP ribose L-system rules**

$$p_1 : C(grow, NAD, t, r) : r \neq \varnothing \longrightarrow gmC(grow, NAD, 0, \varnothing) : 91\%$$
$$p_2 : C(grow, NAD, t, r) : r \neq \varnothing \longrightarrow gmC(branch, NAD, 0, \varnothing)C(grow, NAD, 0, \varnothing) : 9\%$$
$$p_3 : C(branch, NAD, t, r) : r \neq \varnothing \longrightarrow bmC(grow, NAD, 0, \varnothing)$$
$$p_4 : C(grow, NAD, t, r) : t > t_{max} \longrightarrow \varepsilon$$
$$p_5 : C(branch, NAD, t, r) : t > t_{max} \longrightarrow \varepsilon$$

**Appendix 3: Microtubules L-system rules**

$$p_1 : C(grow, tubulin, t, r) : r \neq \varnothing \longrightarrow vbhaC(grow, tubulin, 0, \varnothing)$$
$$p_2 : C(grow, tubulin, t, r) : t > t_{max} \longrightarrow \varepsilon$$

**Appendix 4: Synthetic L-system rules**

$$\begin{aligned}
&p_1 : C(aGrow, a, t, r) : r \neq \varnothing \longrightarrow a_g aC(aGrow, a, 0, \varnothing)90\% \\
&p_2 : C(aGrow, a, t, r) : r \neq \varnothing \longrightarrow a_g a \\
&\qquad C(aBranch, a, 0, \varnothing) \\
&\qquad C(aGrow, a, 0, \varnothing)9\% \\
&p_3 : C(aGrow, a, t, r) : r \neq \varnothing \longrightarrow a_g a \\
&\qquad C(aStar1, b, 0, \varnothing) \\
&\qquad C(aStar2, b, 0, \varnothing) \\
&\qquad C(aStar3, b, 0, \varnothing)1\% \\
&p_4 : C(aBranch, a, t, r) : r \neq \varnothing \longrightarrow a_b aC(aBranchGrow, a, 0, \varnothing) \\
&p_5 : C(aBranchGrow, a, t, r) : r \neq \varnothing \longrightarrow a_g aC(aBranchGrow, a, 0, \varnothing) \\
&p_6 : C(aStar1, b, t, r) : r \neq \varnothing \longrightarrow a_{s1} bC(bGrow, c, 0, \varnothing) \\
&p_7 : C(aStar2, b, t, r) : r \neq \varnothing \longrightarrow a_{s2} bC(bGrow, c, 0, \varnothing) \\
&p_8 : C(aStar3, b, t, r) : r \neq \varnothing \longrightarrow a_{s3} bC(bGrow, c, 0, \varnothing) \\
&p_9 : C(bGrow, c, t, r) : r \neq \varnothing \longrightarrow b_g cC(cGrow, b, 0, \varnothing) \\
&p_{10} : C(cGrow, b, t, r) : r \neq \varnothing \longrightarrow c_g bC(bGrow, c, 0, \varnothing)
\end{aligned}$$

**Author details**

[1] Department of Informatics, University of Bergen, N-5020 Bergen, Norway. [2] The Institute of Computer Graphics and Algorithms, Vienna University of Technology, Wien, Austria.

**Figures**

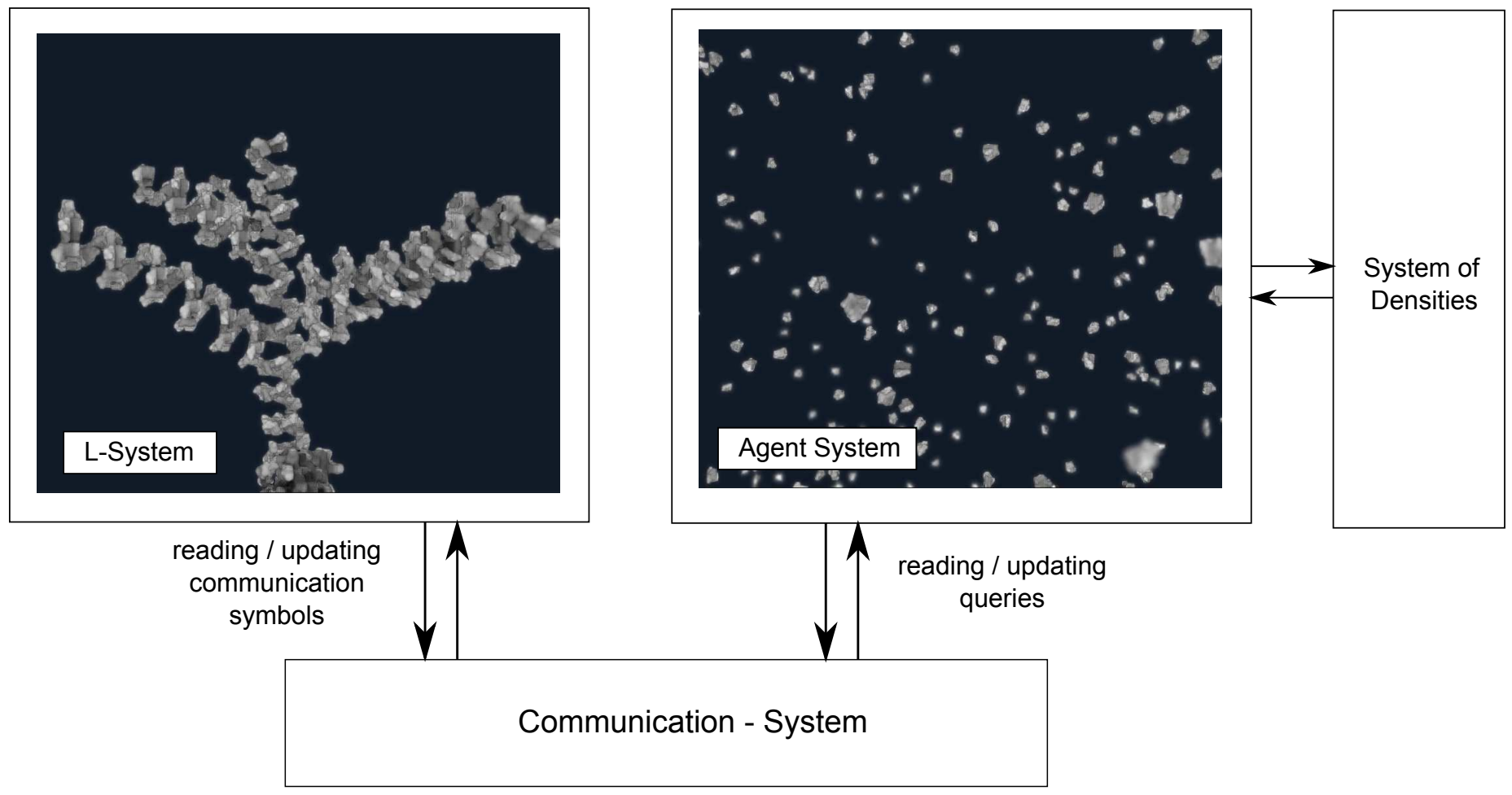

**Figure 1** The overview of our system. An overall process is controlled by the L-System. Communication between L-System and Agent System is performed through the communication symbol that, using the Communication system, is transferred to the Agent System in the form of queries. Results from the queries are written back to the communication symbols and processed by L-System's production rules. The system of densities provides means to change the amount of agents in the environment.

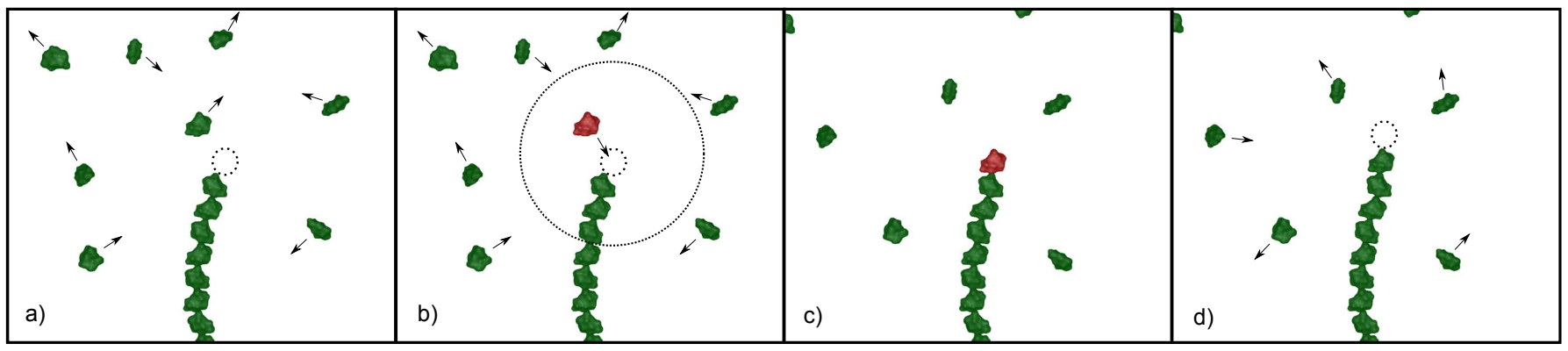

**Figure 2** Illustration of several simulation steps. After the initial configuration (a), a communication symbol was generated, which attracts nearby agents (b). When the agent arrives at the binding site, it is attached to the structure and the communication symbol is terminated (c). Afterwards a new communication symbol is created by production rules and is again attracting nearby agents (d).

**Tables**

**Table 1** Selected strengths and weaknesses of L-Systems vs. Agent-based systems

| Modeling approach | Strengths | Weaknesses |
|---|---|---|
| **L-systems** | Suitable for modeling structures from empirical knowledge. | Limitations writing the creation of structure from stochastically behaved individual entities. |
| **Agent-based systems** | Ability to simulate a stochastic environment. | The global effect, resulting from the interaction of the individuals, is quite unpredictable. |

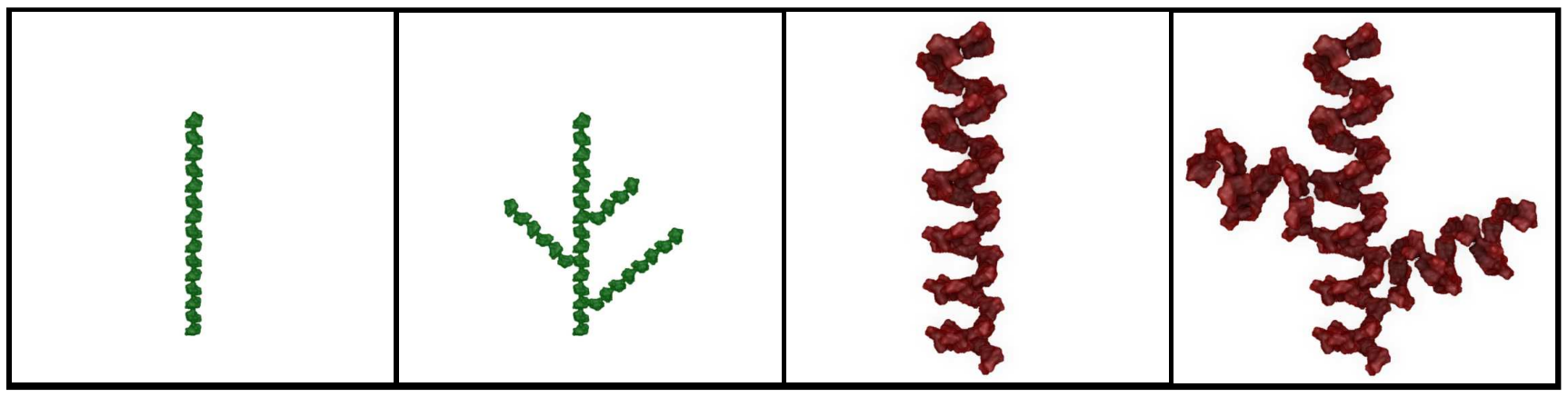

**Figure 3** Illustration of L-system interpretation for linear (a, c) and branched (b, d) monomer structures with a straight (a,b) and spiral (c, d) branch shapes.

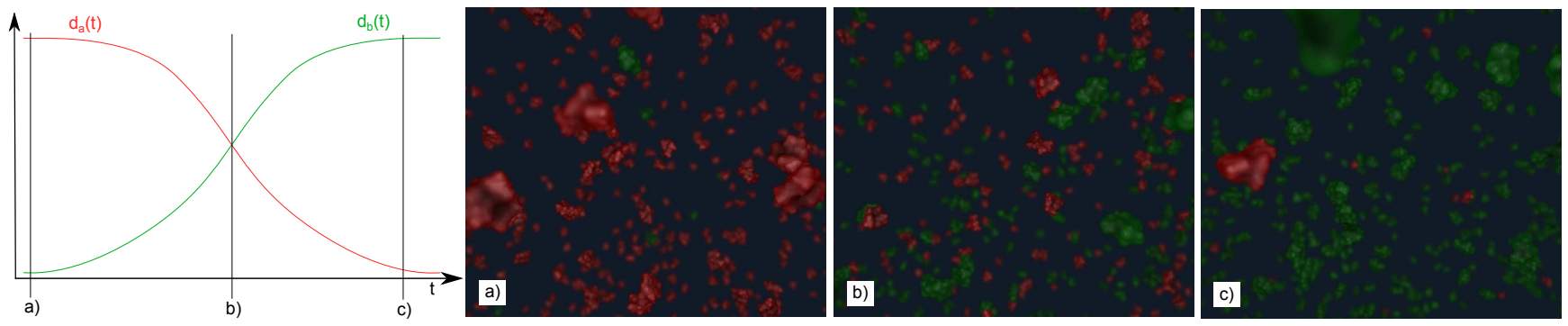

**Figure 4** An example of density-based steering possibilities for the agent-based system. Two agent density functions change the number of agents in the agent-based system over time (from left to right).

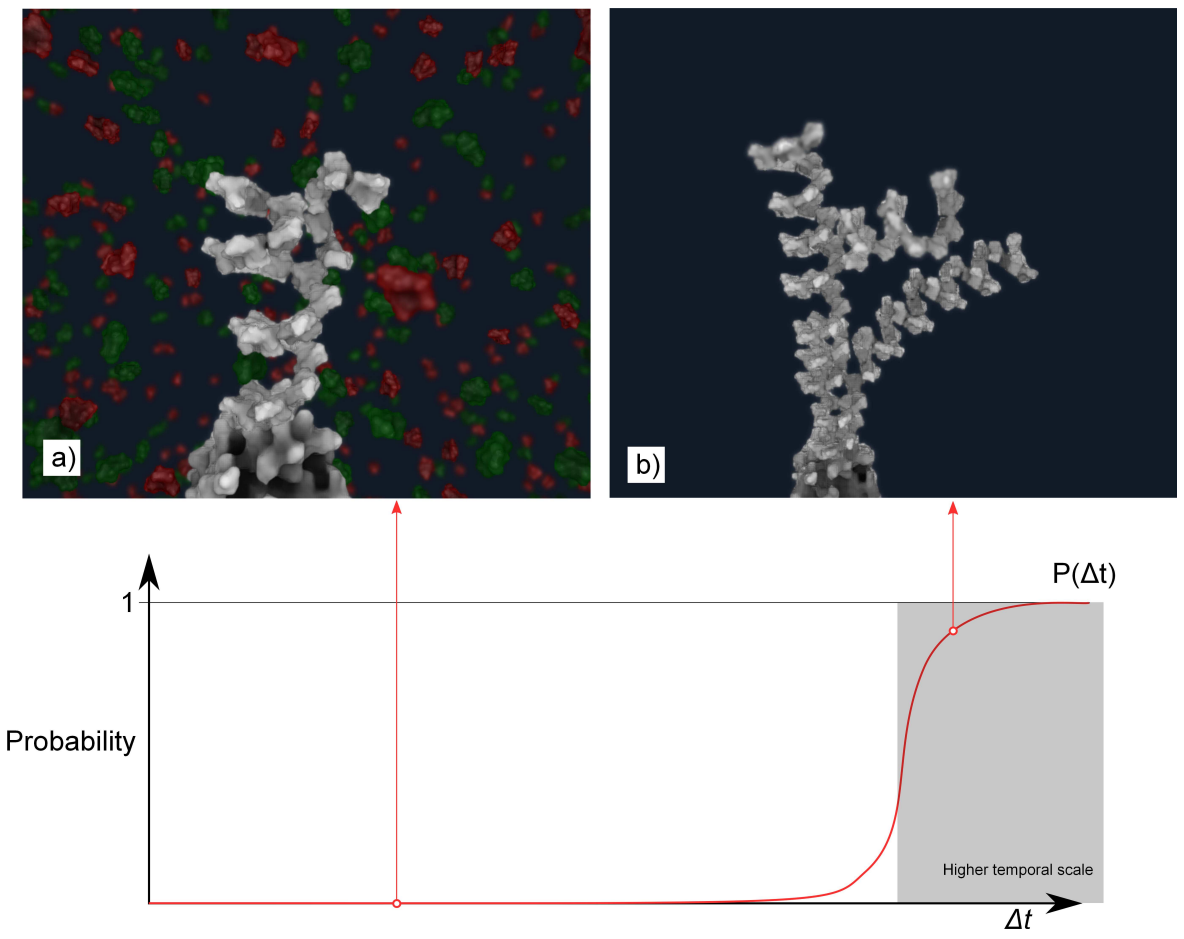

**Figure 5** An example of the probability function $P(\Delta t)$. When $P(\Delta t) = 0$ the simulation runs in the lower time scale (a), while for higher values it runs in the time scale of the entire process (b).

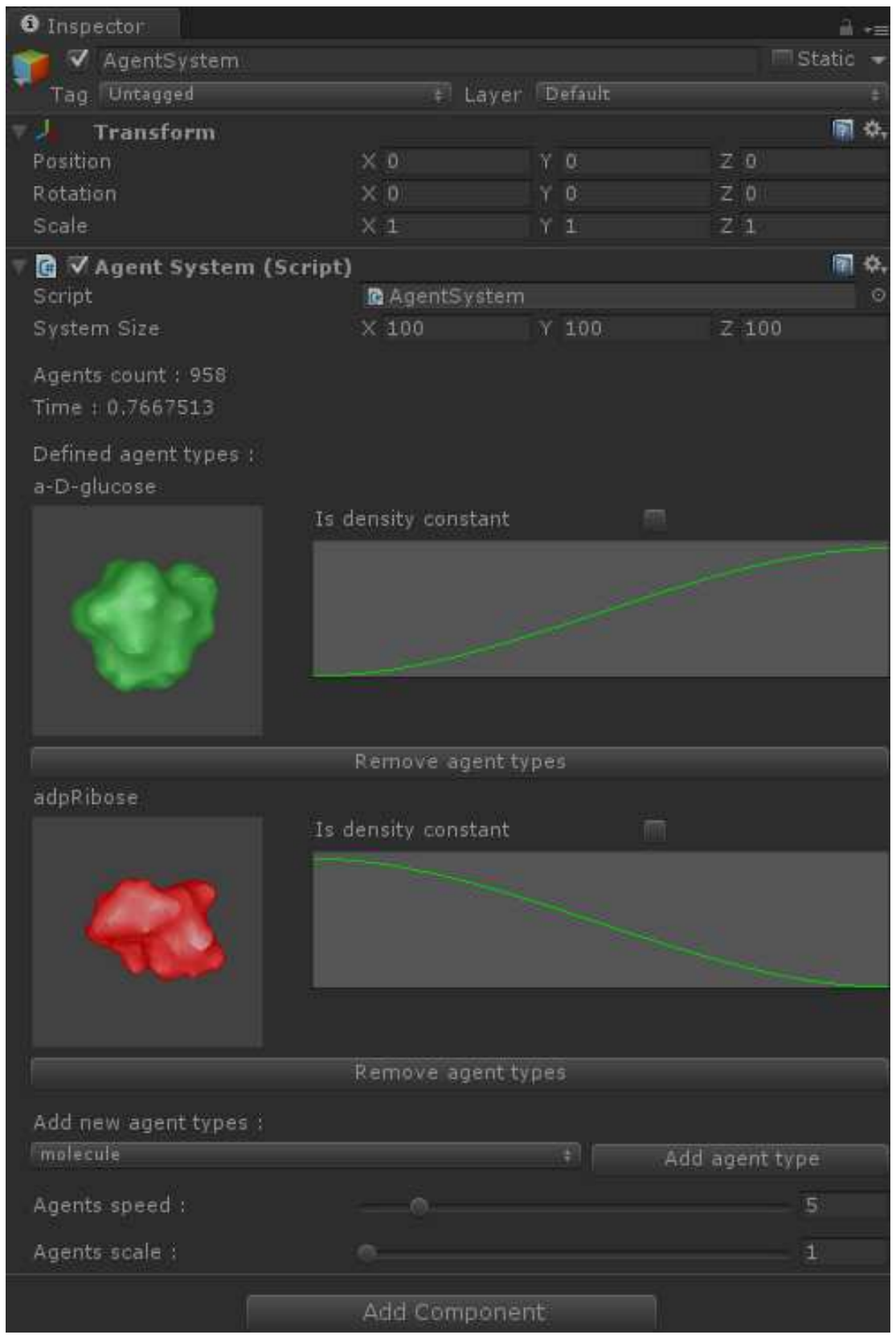

**Figure 6** An application screenshot depicting the Unity3D GUI for editing agents and their densities.

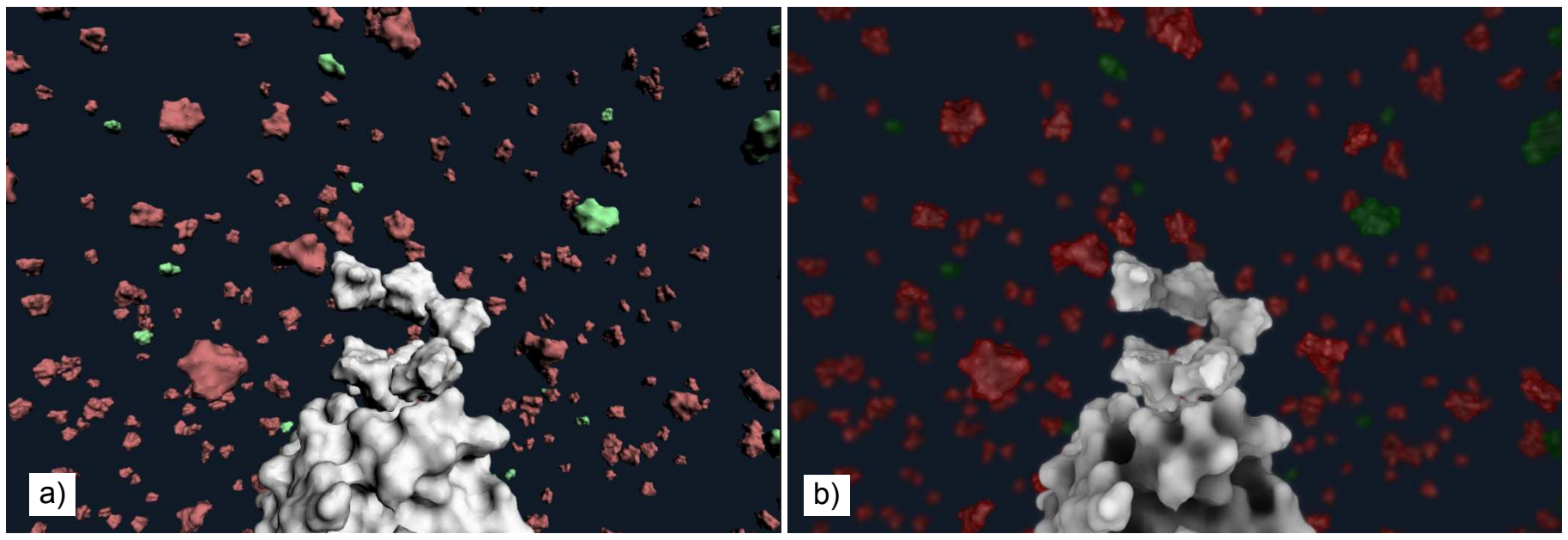

**Figure 7** A comparison between basic Unity3D diffuse rendering (a) and the additional use of screen space effects (b).

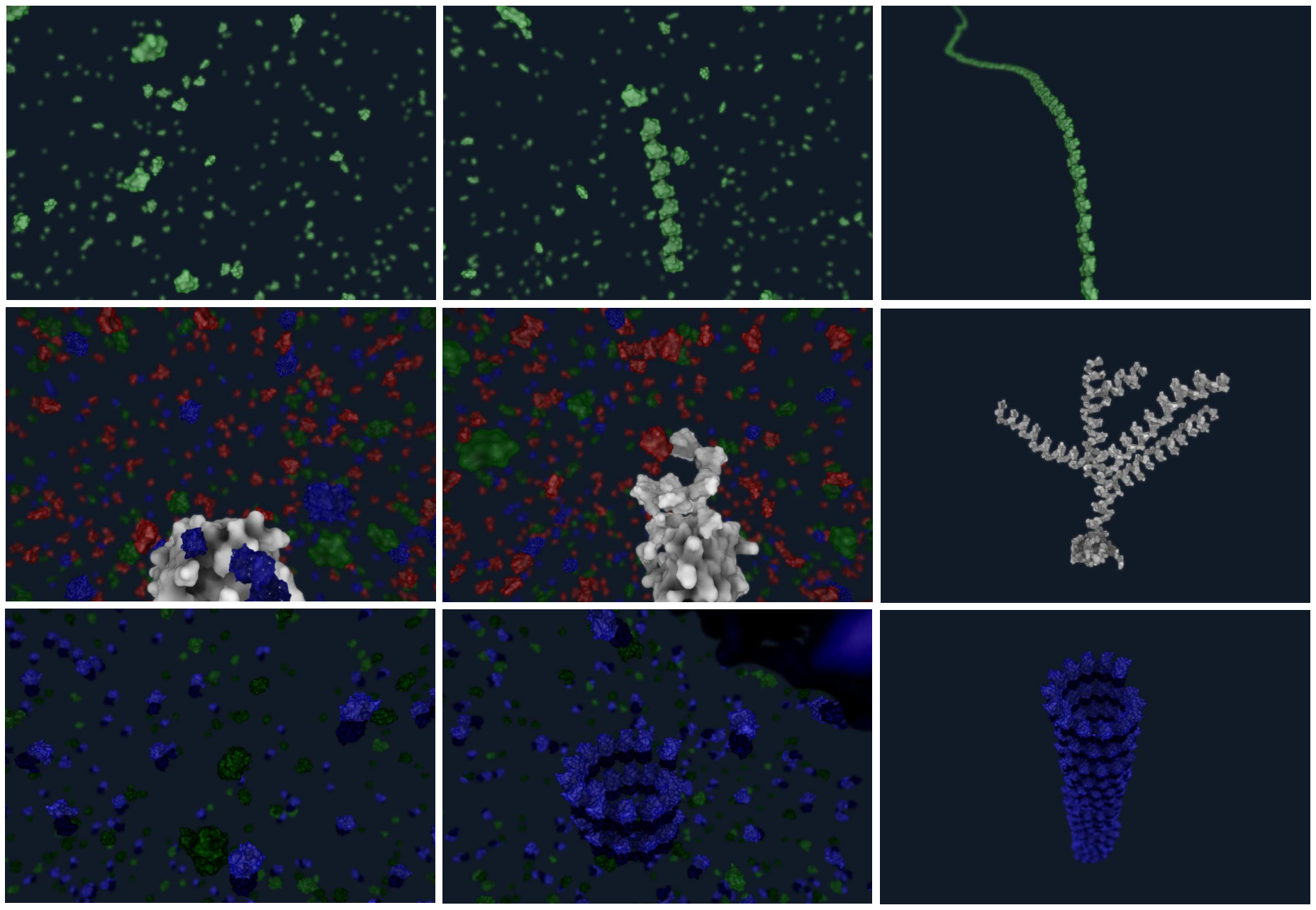

**Figure 8** An example of three polymerization processes: cellulose (1. row), poly-ADP ribose (2. rows) and microtubule (3.row). Column-wise, the ordering (from left to right) represents the start of simulation, illustration of processes, and final structure.

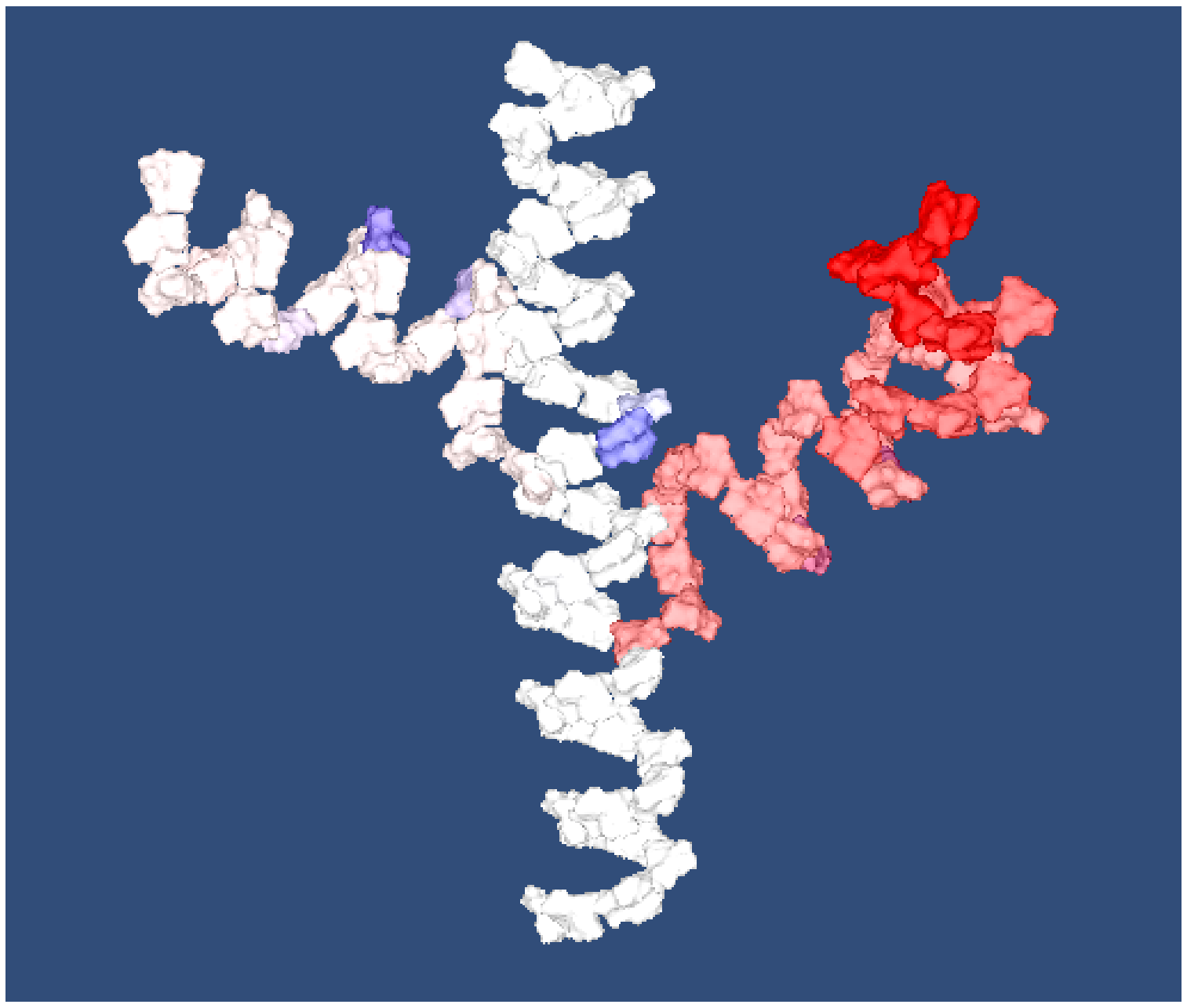

**Figure 9** An visualization example of branching uncertainty and branching probability factor in the resulting structure. Branching uncertainty is accumulated using a Gaussian kernel centered at the probability threshold for branching and growing. It is visualized in the new branches as transition from white, no uncertainty, to red, high uncertainty. The blue color indicates how the branching probability for cases when no new branch was created.